\documentclass{elsart3p}

\usepackage{graphicx,color}
\usepackage{amssymb}
\begin{document}
\begin{frontmatter}
%titles, authors and addresses

\title{Superconducting ground state of the two-dimensional Hubbard model: a variational study}


\author{David Eichenberger and Dionys Baeriswyl}

\address{D\'epartement de physique, Universit\'e de Fribourg, Chemin du mus\'ee 3, CH-1700 Fribourg, Switzerland}

\maketitle

\begin{abstract}
A trial wave function is proposed for studying the instability of the two-dimensional Hubbard 
model with respect to $d$-wave superconductivity. Double occupancy is reduced in a similar
way as in previous variational studies, but in addition our wave function both enhances the
delocalization of holes and induces a kinetic exchange between the electron spins. These
refinements lead to a large energy gain, while the pairing appears to be weakly affected by the additional term in the variational wave function.
\end{abstract}
%\begin{keyword}
% keywords here, in the form: keyword; keyword
%Hubbard model\sep variational method 

%\PACS 
%\end{keyword}
\end{frontmatter}

% main text
\section{Introduction}
\label{Intro}

The insulating antiferromagnetic phase of layered cuprates is well described by the 
two-dimensional Hubbard model at half filling. It is less clear whether this model is
also able to describe the superconducting phase observed for hole concentrations 
$0.25<p<0.5$. It has been shown a long time ago that fermions with purely repulsive
interactions can become superconducting, but the initial estimates for the 
critical temperature of continum models were deceptively low \cite{Kohn}. In the mean-time, both analytical and numerical studies have indicated that for electrons on a lattice the situation may not be hopeless. Unfortunately, reliable estimates for the superconducting gap $\Delta$, the condensation energy $W_n-W_s$ or other important quantities for the two-dimensional repulsive Hubbard model and variants thereof are still missing.

Here we report on the current status of our variational studies of the two-dimensional
Hubbard model for intermediate values of $U$ and a hole doping $p\approx 0.19$. Our
preliminary results for the gap and the condensation energy are consistent with typical 
experimental values.

\section{Variational approach}

There are two competing terms in the Hubbard Hamiltonian $\hat{H}=-t\hat{T}+U\hat{D}$, 
the hopping term between nearest-neighbour sites

\begin{equation}
\qquad\qquad\hat{T}=\sum_{\langle i,j\rangle,\sigma}(c^\dag_{i\sigma}c_{j\sigma}+c^\dag_{j\sigma}c_{i\sigma})\, ,
\end{equation}
where $c^\dag_{i\sigma}$ creates an electron at site $i$ with spin $\sigma$, and the on-site interaction (the number of doubly occupied sites)

\begin{equation} 
\qquad\qquad\hat{D}=\sum_i n_{i\uparrow}n_{i\downarrow}\, ,\quad n_{i\sigma}=c^\dag_{i\sigma}c_{i\sigma}\, .
\end{equation}

 Several variational studies have been performed for the limiting case $U\rightarrow\infty$,
where double occupancy is completely suppressed \cite{Gros,Paramekanti}. This limit is
not appropriate for intermediate values of $U$. We have thus chosen the variational ansatz  

\begin{equation}
\label{wf_def}
\qquad\qquad\vert\Phi\rangle=e^{-h\hat T}e^{-g\hat D}\vert d\mbox{BCS}\rangle\, ,
\end{equation}
where double occupancy is only partially suppressed (variational parameter $g$). At the same
time both the delocalization of holes and kinetic exchange between spins are enhanced (parameter $h$). The parent state $\vert d\mbox{BCS}\rangle$ is a BCS state with parameters describing $d$-wave pairing, {\it i.e.,}  
\begin{eqnarray}
\qquad u_{\bf{k}}^2=\frac{1}{2}(1+\frac{\epsilon_{\bf{k}}-\mu}{E_{\bf{k}}}),\qquad u_{\bf{k}}v_{\bf{k}}=\frac{\Delta_{\bf{k}}}{2E_{\bf{k}}}\nonumber
\end{eqnarray}
with
\begin{eqnarray}
E_{\bf{k}}&=&\sqrt{(\epsilon_{\bf{k}}-\mu)^2+\Delta_{\bf{k}}^2},\ \Delta_{\bf{k}}=\Delta\cdot(\cos{k_x}-\cos{k_y}).\nonumber
\end{eqnarray}
We notice that our wave function contains two additional variational parameters, the gap $\Delta$ and the ``chemical potential'' $\mu$ ($\mu$ is a variational parameter and not the true chemical potential). 

The expectation value of the Hamiltonian with respect to our trial state is computed using a Monte Carlo simulation. This is straightforward for $h=0$, where the BCS state can be projected onto a subspace with a fixed number of particles and written as a superposition of real space configurations. For $h>0$ the projection onto a state with a fixed number of particles is found to lead to minus sign problems, which worsen as the gap parameter increases. The problem can be solved by using a fixed ``chemical potential'' (instead of a fixed number of particles) together with a momentum space representation. A Hubbard-Stratonovich transformation is used to decouple the on-site interaction in the Gutzwiller projector. Unfortunately, this approach results in a very slow convergence.

\begin{figure}[h]
\hspace{0.2cm}
\includegraphics[width=0.42\textwidth]{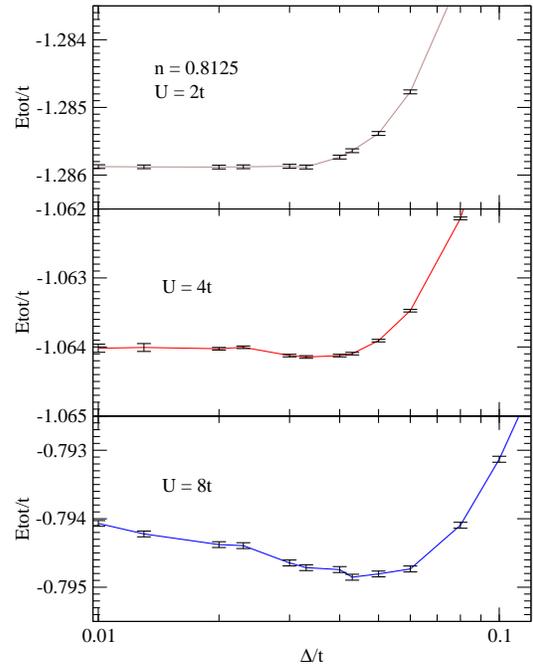}
\caption{\label{Gutz}Total energy per site of a 8x8 square lattice for a density corresponding to the slightly overdoped region of the phase diagram of cuprates.}
\end{figure}

\section{Results and conclusions}
\label{result}

We have first considered the particular case $h=0$, which has been studied previously
\cite{Giamarchi}. Fig.\  \ref{Gutz} shows the energy per site as a function of the gap $\Delta$ for several values of $U$. For $U=2t$ the numerical precision does not allow to draw
any conclusion about pairing, but for $U=4t$ and $U=8t$ there are clear minima at
$\Delta \approx 0.04t $ and $\Delta\approx 0.05t$, respectively.

Allowing $h$ to vary improves significantly both the energy and the wave function
\cite{Otsuka,Dzier}. For $U=8t$, the energy gain due to the refinement of the variational ansatz is two orders of magnitude larger than the condensation energy of Fig.\ \ref{Gutz}. Nevertheless, our first results  for $h>0$ confirm the trend towards $d$-wave superconductivity, with a gap of about the same size as in Fig.\ \ref{Gutz}.

We briefly comment on the comparison between our results (for $U=8t$, $t=300$ meV and $n=1-p\approx 0.81$) and experimental data for the layered cuprates. Typical data for the gap parameter found in photoemission experiments are $\Delta\approx 10-15$ meV \cite{shen}, while the condensation energy obtained from specific heat data is of the order of $0.1$ meV \cite{Loram}. Both experimental values agree surprisingly well with our variational results.

\vspace{0.28cm}
\begin{center}
{\bf Acknowledgement}
\end{center}
\vspace{0.28cm}

This work was supported by the Swiss National Science Foundation through the National Center of Competence in Research ``Materials with Novel Electronic Properties-MaNEP''.

\end{document}